\documentclass[floatfix,aip,amsmath,amssymb,reprint]{revtex4-1}
 
\usepackage{graphicx}
\usepackage{dcolumn}
\usepackage{bm}

\usepackage[utf8]{inputenc}
\usepackage[T1]{fontenc}
\usepackage{mathptmx}
\usepackage{etoolbox}
\usepackage[dvipsnames]{xcolor}
\makeatletter
\def\@email#1#2{%
 \endgroup
 \patchcmd{\titleblock@produce}
  {\frontmatter@RRAPformat}
  {\frontmatter@RRAPformat{\produce@RRAP{*#1\href{mailto:#2}{#2}}}\frontmatter@RRAPformat}
  {}{}
}%
\makeatother
\begin{document}

\preprint{AIP/123-QED}

\title[Magnetic field evolution in jets]{Comprehensive study of magnetic field evolution in relativistic jets based on 2D simulations}
\author{Amin Esmaeili }
\affiliation{Department of Physics, Graduate School of Science, Tokyo Metropolitan University, 1-1 Minami-Osawa, Hachioji-shi, Tokyo 192-0397, Japan}
\email{amin.esmaeili@gmail.com}
\author{Yutaka Fujita}
\affiliation{Department of Physics, Graduate School of Science, Tokyo Metropolitan University, 1-1 Minami-Osawa, Hachioji-shi, Tokyo 192-0397, Japan}

\date{\today}

\begin{abstract}
We use two-dimensional particle-in-cell simulations to investigate the generation and evolution of the magnetic field associated with the propagation of a jet for various initial conditions. We demonstrate that, in general, the magnetic field is initially grown by the Weibel and Mushroom instabilities. However, the field is saturated by the Alfv\'en current limit.
For initially non-magnetized plasma, we show that the growth of the magnetic field is delayed when the matter density of the jet environment is lower, which are in agreement with simple analytical predictions. We show that the higher Lorentz factor ($\gtrsim 2$) prevents rapid growth of the magnetic fields. When the initial field is troidal, the position of the magnetic filaments moves away from the jet as the field strength increases. The axial initial field helps the jet maintain its shape more effectively than the troidal initial field.
\end{abstract}

\maketitle

\section{\label{sec:level1}Introduction:}

Relativistic jets are often found in astrophysical objects such as pulsars, gamma ray bursts
(GRB), and active galactic nuclei (AGN).  They are magnetized because synchrotron radiation has been detected\cite{gabuzda19}.
When a relativistic jet interacts with its environment, several instabilities grow. These may be responsible for the growth of the magnetic field. For example, the Weibel instability has been shown to develop in non-magnetized jets \cite{weibel59}. 
This instability occurs when the particle velocity distribution function (PDF) is anisotropic. The Weibel instability helps to make the PDF isotropic by generating magnetic fields and currents in the plasma to deflect the plasma particles \cite{fujita06}. The shear of the jet boundary also causes
instabilities such as the kinetic Kelvin-Helmholtz instability (kKHi) and the mushroom
instability (MI); the MI develops perpendicular to the direction of the jet \cite{rieger21, meli21, alves18}. We note that the MI instability is a Rayleigh-Taylor instability (RTi) caused by electrons/ions on the kinetic scale. 
After the initial development, the magnetic field gradually saturates, which has been studied for simple settings \cite{kato05, fujita06}.

The method of simulating plasma phenomena using moving particles was developed by
a group of scientists, including Buneman, Hockney, Birdsall, and Dawson. In this method or particle-in-cell (PIC) simulations, the particles are controlled by the electromagnetic field produced by their own movement and by any external fields. The electromagnetic
field alters the particle's trajectory in a self-consistent manner. The method does not analyze bulk/fluid
dynamics like the magnetohydrodynamics (MHD) code, but instead takes a kinematic approach to the problem and 
complements the fluid approach. However, it can be challenging for the PIC simulations to handle a wide range of scales, spanning from microphysics to macro/global dynamics, although it is necessary to study AGN jets \cite{nishikawa21}.

In this letter, we present the results of two-dimensional (2D) PIC simulations that reveal magnetic field generation and its impact on jet evolution. We study the evolution of the field in the cross-section of the
jet, which is an analysis unfeasible through traditional fluid-based approaches. 
We focus on a numerical study of basic plasma phenomena rather than possible applications to AGN jets, since current computing power does not allow us to reproduce a realistic AGN jet on the scale of $>$~pc using PIC simulations.
We run simulations with different setups of magnetized and unmagnetized jets. 
Our simulations are similar to those of Kawashima et al.
\cite{kawashima22}, where they focused mainly on the effect of magnetic reconnection on the heating of the jet. They showed that the x-point of the magnetic reconnection moves inward over time, which helps to maintain the spine of the jet. 
On the contrary, we study the effects of different magnetic field structures on the jet propagation and the instability developed along the jet. 
Since our main goal is to reveal the parameter dependence of plasma instabilities, we need to cover a wide range of parameters. Therefore, we perform 2D PIC simulations instead of 3D. We also note that we can discuss the $z$ components of the magnetic field and velocity even though our 2D simulations deal with the $xy$ cross section of the jet.

\section{\label{sec:level1-1}simulation setup:}

Our code has two spatial and three
velocity dimensions (2D3V). The simulation plane
is taken on the $xy$ plane perpendicular to the $z$-axis.
The jet propagates in the $z$-direction.
We adopt a simulation setup similar to that of Kawashima et al.\cite{kawashima22}, and the parameters are shown in Table~\ref{tab:table1}.
The ion mass $m_i$ to electron mass $m_e$ ratio is $m_i/m_e=16$ in most runs and the thermal velocity of the plasma particles in the $x$, $y$ and $z$ directions is $v_{\rm thr}=0.1\: c$, where $c$ is the light velocity. 
We used this ion-electron mass ratio of $m_i/m_e=16$ because a realistic mass ratio requires more time and space resolution, and it is difficult to cover a wide range of initial parameters. However, we perform simulations of $m_i/m_e=1836$ for a few representative cases to check consistency (section E.3).
Different initial bulk jet velocities $0.25\: c \leq V_{c0} \leq 0.948\: c$ are applied in the $z$-direction. The corresponding Lorentz factor is given by $\gamma_0=1/\sqrt{1-V_{c0}^2/c^2}$.
Since we also studied the effects of different jet environments as in Alves et al.\cite{alves14}, we assume different ratios of the initial density of the jet ($n_{\rm jet}$) to that of the surrounding environment ($n_{\rm out}$). We set the number density of the jet plasma to $n_{\rm jet}=100$.
We consider zero, troidal, and axial fields for the initial magnetic field, and the characteristic value is given by $b_0$. Even when $b_0=0$, magnetic fields are generated around the jet boundary. When the initial magnetic field is toroidal, the $x$ and $y$ components are given by $b_{x0}=(y-y_{jc})b_{0}/[1+(r/a)^{2}]$, and $b_{y0}=(x-x_{jc})b_{0}/[1+(r/a)^{2}]$, respectively, where $(x_{jc}, y_{jc})$ is the jet axis and $a$ is the jet radius. For models with a zero or axial initial magnetic field, the field strength is spatially uniform and is given by $b_0$.

The grid sizes are $\Delta x=\Delta y=1$, plasma frequency is $\omega_{pe}=0.05$, Debye length is
$\lambda_{D}=0.05$ , skin depth is $\lambda_{s}=c/ \omega_{\rm pe}=20$, simulation time-step is $dt=0.5$, simulation box size is $800 \times 800$, and total
simulation time is $t=500$. The time unit is $\omega_{pe}^{-1}$. 
The jet axis is at $(x_{jc}, y_{jc}) = (400, 400)$ except for Run~11 and the initial jet radius is $a =100$.
We use 100 electron-ion pairs in
each cell. Periodic boundary
conditions are adopted in the $x$ and $y$-directions for both particles and electromagnetic
fields. To verify convergence, similar runs were performed with more particles in each cell (150 and 200 particles), and we confirmed that the results converged as the number of particles in each cell increased.

We note that the width of real astrophysical jets is $\gtrsim 10^{16}\rm\: cm$,\cite{2011Natur}
while the phyical diameter of the jet used in our simulation is about $10\: c/\omega_{pe}\sim 10^5\rm\: cm$, (if we assume the number density of the jet to be about $10^4\rm\: cm^{-3}$)\cite{kawashima22}. Thus, our length scale is much smaller compared to the real jet.

\begin{table}
\caption{\label{tab:table1}Simulation Parameters}
\begin{ruledtabular}
\begin{tabular}{ccccccccc}
 Run &Initial magnetic field &$V_{c0}$ &$\gamma_0$ &$n_{\rm out}$ &$n_{\rm jet} /n_{\rm out}$\\
\hline
1& $b_0=0.0$ & $0.9c$ &2.3 &100 &1\\
5& Toroidal, $b_0=0.01$\footnotemark[1] & $0.9c$ &2.3 &100 &1\\
6& Toroidal, $b_0=0.05$\footnotemark[1] & $0.9c$ &2.3 &100 &1\\
8& Toroidal, $b_0=0.01$\footnotemark[1] & $0.5\rm c$ &1.16 &100 &1\\
9& Toroidal, $b_0=0.01$\footnotemark[1] & $0.25\rm c$ &1.033 &100 &1\\
10& Toroidal, $b_0=0.025$\footnotemark[1] & $0.5\rm c$ &1.16 &100 &1\\
11& Toroidal, $b_0=0.01$\footnotemark[2] & $0.5\rm c$ &1.16 &100 &1\\
12& Axial, $b_{0}=0.01$ & $0.5\rm c$ &1.16 &100 &1\\
13& Axial, $b_{0}=0.025$ & $0.5\rm c$ &1.16 &100 &1\\
17& Axial, $b_{0}=0.01$ & $0.9\rm c$ &2.3 &100 &1\\
30& $b_0=0.0$ & $0.5\rm c$ &1.16  &100 &1\\
31& $b_0=0.0$ & $0.5\rm c$ &1.16  &80 &1.25\\
32& $b_0=0.0$ & $0.5\rm c$ &1.16  &60 &1.66\\
33& $b_0=0.0$ & $0.5\rm c$ &1.16  &40 &2.5\\
34& $b_0=0.0$ & $0.5\rm c$ &1.16  &20 &5.0\\
36& $b_0=0.0$ & $0.866\rm c$ &2 &100 &1\\
37& $b_0=0.0$ & $0.948\rm c$ &10 &100 &1\\
\end{tabular}
\end{ruledtabular}
\footnotetext[1]{$b_{x0}=(y-y_{jc})b_{0}/[1+(r/a)^{2}]$ and $b_{y0}=(x-x_{jc})b_{0}/[1+(r/a)^{2}]$, where $(x_{jc}, y_{jc}) = (400, 400)$.}
\footnotetext[2]{$b_{x0}=(y-y_{jc})b_{0}/[1+(r/a)^{2}]$ and $b_{y0}=(x-x_{jc})b_{0}/[1+(r/a)^{2}]$, where $(x_{jc}, y_{jc}) = (420, 420)$.}
\end{table}

\section{\label{sec:level1_2}simulation results and discussion:}

\subsection{\label{sec:level2}Jet evolution}

\begin{figure*}
\includegraphics[width=7.in]{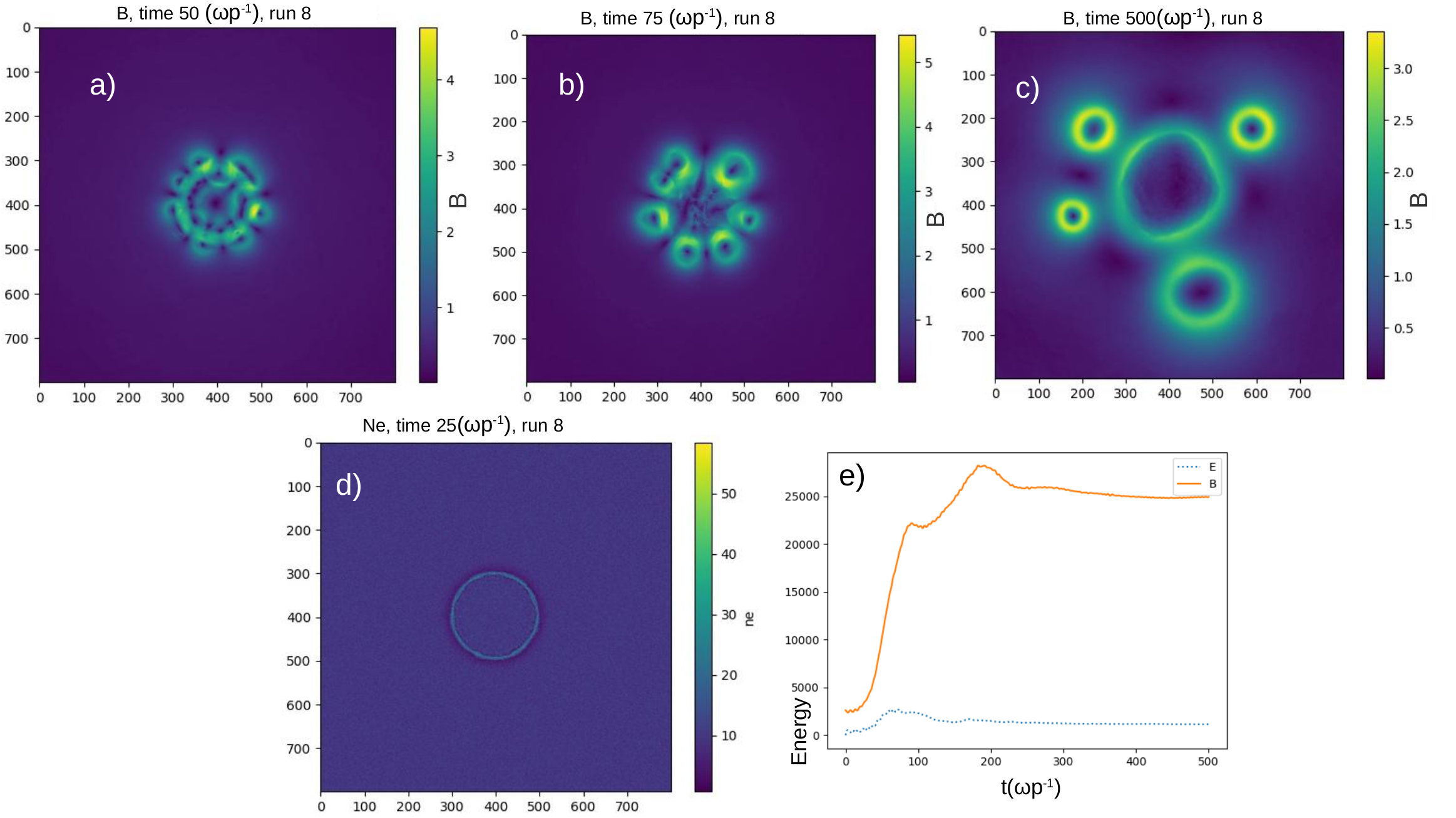}  
\caption{\label{fig1}  
Magnetic field evolution for Run 8 ($\gamma_0=1.16$ and $b_0=0.01$). The initial jet
velocity is $0.5\: c$ normal to the paper. (a) Magnetic field strength at $t=50$, (b) $t=75$, and (c) $t=500$. In panel (a) the field is generated and in panels (b), (c) and (d) the merging of filaments at the jet boundary is seen. (d) Electron number density at the time $t=25$. Electrons are slightly clustered around the jet. (e) Evolution of total magnetic energy (orange solid line) and electric energy (cyan dotted line).}
\end{figure*}

We show overall jet evolution by taking Run 8 as the fiducial model (Figure~\ref{fig1}).
Initially, particles in the simulation box have a bulk velocity in the $+z$ direction. Therefore, electrons
carry current in the $-z$ direction and ions carry current in the opposite direction. As a
result of the anisotropy of the PDF, filaments are formed in the $z$-direction by the Weibel instability; the main filaments are formed in the jet radius and other filaments are formed in the surrounding region (Figure~\ref{fig1}(a)). The rings shown in Figures~\ref{fig1}(a)-(c) are the cross-section of the filaments. Magnetic fields are generated around the currents associated with the filaments. The filaments attract each other and merge together by magnetic force (Figures~\ref{fig1}(b) and (c)), which increases the current.  Due to the amplification of the magnetic field, the particles become more deflected. 
Both the current and
the magnetic field continue to grow exponentially through the mergers of filaments until saturation is
reached. The current tubes merge until $t\sim 200$. Then they stop growing. The magnetic field also saturates around that time  (Figure~\ref{fig1}(e)).

Figure \ref{fig1}(d) shows the electron density distribution at time $t=25$, which is almost uniform but slightly clustered around the jet.
After the saturation ($t\sim 200$), the size of the current tubes remains almost the same.
There is a small saturation before $t\sim 100$ (Figure \ref{fig1}(e)), which is due to the saturation of the magnetic field
structures inside the jet.

\subsection{Saturation mechanism}

The saturation we showed above may be explained by the model of Kato\cite{kato05}. 
A cylindrical beam is characterized by its radius $r$ and current $I$, where $I=\pi r^{2}J$ and $J$ is the current density. The average magnetic field within the beam can be written as $B=\sqrt{2}\pi rJc=\sqrt{2}\pi I/(rc)$. 
Due to the coalescence of the filaments with each other, both the current and the magnetic field of the filaments increase exponentially, and the filament radius also increases. 
However, this growth stops when the current reaches the Alfven current $I_{A}=\left\langle I_{0} \gamma_0 \beta_{\parallel } \right\rangle$, where  $I_{0}=m_{e}c^{3}e$, $-e$ is the electron charge. $\beta_{\parallel }$ is the $z$ component of $\beta$ (velocity normalized by $c$), $\gamma = (1-\beta^2)^{-1/2}$, and $\left\langle \right\rangle$ denotes averaging over the filament volume. This is because the movement of the particles in the current is impeded by the self-generated magnetic field.

From the Alfven current, the maximum magnetic field is calculated as: 
\begin{equation}
\label{eq:Bmax}
    B_{\rm max}=\frac{\sqrt{2}}{rc}I_{\rm A}=\frac{\sqrt{2}\left\langle \gamma \beta_{\parallel } \right\rangle}{r/I_{0}}B_{\ast }\:,
\end{equation}
 where $B_{\ast }$ is the magnetic field strength defined by $B_{\ast }=m_{e}c \omega_{\rm pe} e\cong 3.2\times 10^{-1} \sqrt{n_{e}}[\rm G]$.

We compare the maximum or saturated magnetic field around the jet obtained by the simulations  with the theory. Since the model of Kato\cite{kato05} does not include background magnetic fields, we compare it with the results of Run 31, which has the same initial parameters as Run 8 except for the zero initial magnetic field ($b_0=0$); Run 31 shows a similar evolution to Run 8. We find that while the maximum magnetic field in the simulation at time $t=200$ is 2.5, equation~(\ref{eq:Bmax}) gives $B_{\rm max}=2.87$ for $\gamma_0=1.16$, $\beta_\parallel=0.5$, and $I_0=101.58$. Thus, they are close to each other. This is almost the same for Run 8 with non-zero but small $b_0$.

\subsection{Jet and environment density}

For initially non-magnetized plasma, we study the effects of different ambient densities on the jet evolution. First, we discuss analytically the growth rate of the MI.
We use a fluid model of Alves et. al.\cite{alves15} to obtain the growth rate for different density contrasts.
While they only gave results when there was no density contrast, we consider the cases where the densities are different. Here we evaluate the growth rate for the different $n_{\rm jet}/n_{\rm out}$ by considering a linear perturbation theory. 

\begin{figure}  
\includegraphics[width=3.in]{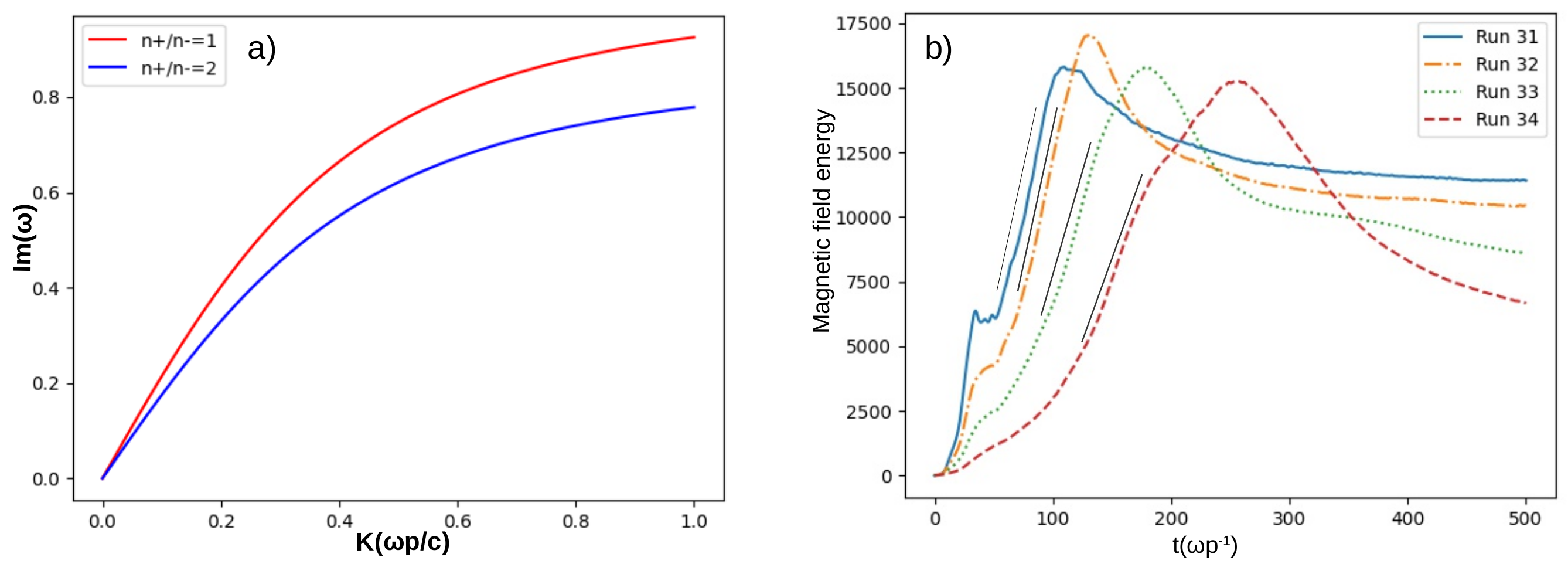}
\caption{\label{fig2} (a) The growth rate of the MI obtained by solving equation~(\ref{eq:det_I}) for  $n_+/n_-=1$ and 2. (b) Simulated evolution of the total magnetic field energy  
for different outer jet densities. The ratio $n_{\rm jet}/n_{\rm out}$ increases from Run~31 to Run~34. The results for Run~30 are almost the same as Run~31. The initial magnetic field is assumed to be 0, and
the initial jet Lorentz factor is assumed to be $\gamma_0=1.16$. The results show that as the outer plasma density decreases, the
saturation time delays and the growth rate decreases (see the gray segments representing the MI growth). There is a slight saturation around time $t=32.5$ due to instabilities in the jet and the saturation of the filaments. }
\end{figure}

\begin{figure*}
\includegraphics[width=7.in]{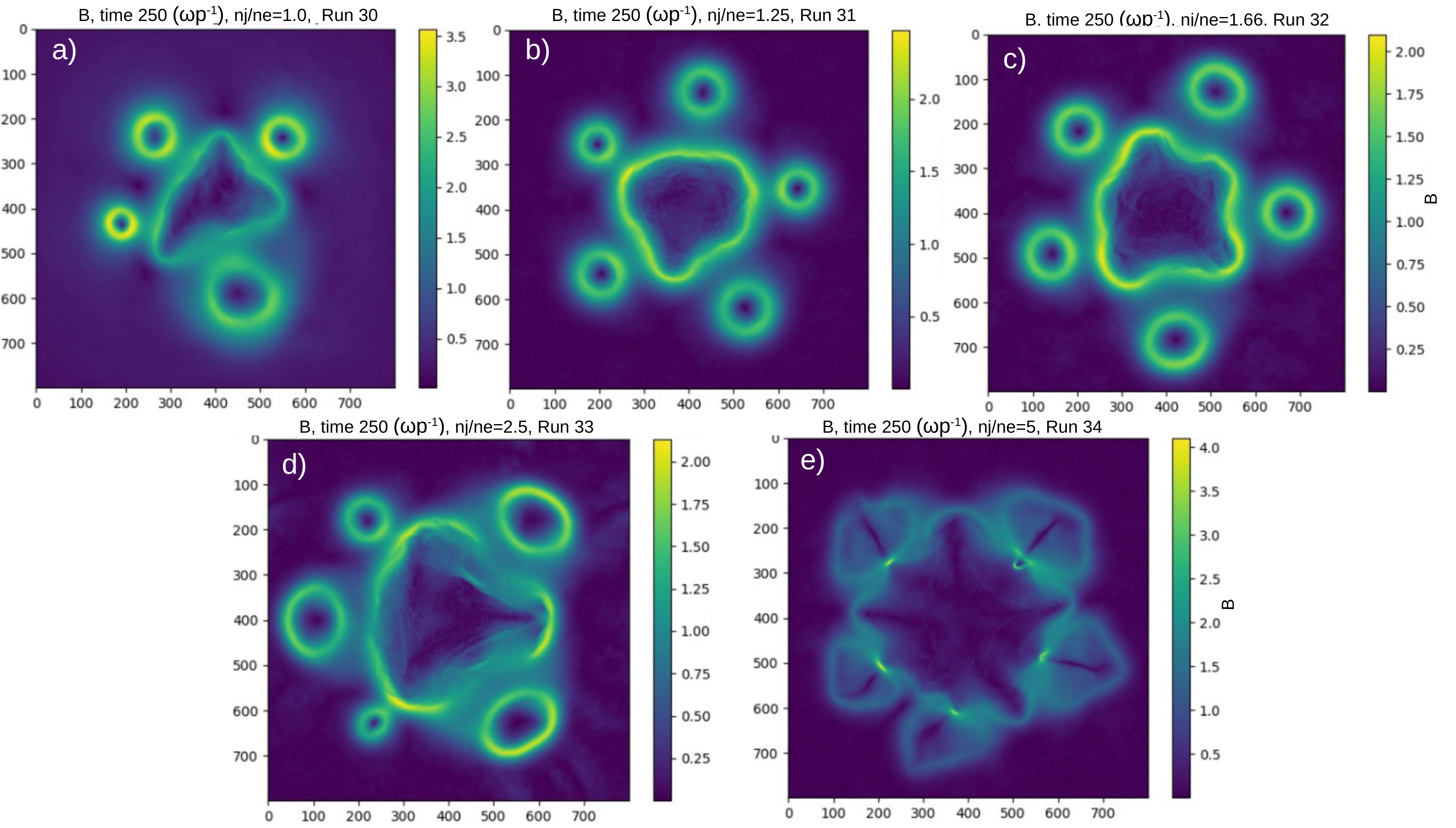}  
\caption{\label{fig2_2} 
Distribution of the magnetic field strength for different $n_{\rm jet}/n_{\rm out}$ at $t\sim 250$ for (a) Run 30, (b) Run 31, (c) Run 32, (d) Run 33, and (e) Run 34 
}\end{figure*}

  We consider two adjacent fluids of different densities; their values are represented by the indices $+$ and $-$, respectively, and their initial or unperturbed values are represented by the index 0. The velocity of the sheared plasma flow is $\vec{v}_0=v_0(x) \vec{e}_z$, where $\vec{e}_z$ is the unit vector in the $z$ direction. 
Considering charge and current neutrality, the equal density ($n_0=n_{i0}=n_{e0}$) and equal velocity ($v_0=v_{i0}=v_{e0}$) conditions are applied to ensure the initial equilibrium between ions ($i$) and electrons ($e$). For a fluid quantity $f$ we assign a perturbation $f=f_0+\delta f$ with $\delta f(x,y,t)=\delta f(x)e^{-i\omega t + iky}$, where $\omega$ ($\in \mathbb{C}$) is the frequency and $k$ ($\in \mathbb{R}$) is the wave number. All zeroth order quantities are assumed to be zero except $n_0(x)$ and $v_0(x)$. 
The perturbed current densities in the $x$, $y$, and $z$ directions are $\delta j_x=-e(1+m_e/m_i)n_0\delta v_{ex}$, $\delta j_y=-e(1+m_e/m_i)n_0\delta v_{ey}$, and $\delta j_z=-e(1+m_e/m_i)(n_0\delta v_{ez}+\delta nv_0)$, respectively. 
A step velocity shear and density profile of the form $v_0(x)=v^-+(v^+-v^-)\mathcal{H}$ and $n_0(x)=n^-+(n^+-n^-)\mathcal{H}$ are used, where $\mathcal{H}$ is a Heaviside step function. 
By integrating $\epsilon.\delta\mathrm{E}=0$ for $x\neq 0$, where $\epsilon_{ij}=\sum_{m=0}^{2}C_{ij,m}[\omega,k_y,v_0(x),n_0(x))]\partial_{x^m}$ ($C_{ij,m}$ is constant factors of equation) and $\delta\mathrm{E}=(\delta E_y,\delta E_z$), and using the continuity of $\delta E_y$ and $\delta E_z$ at the shear interface, we discover solutions that correspond to evanescent waves known as $ \delta E_{y,z} = \delta E_{y,z}(0) e^{-k^{\pm}_{\bot }|x|}$. $k^{\pm}_{\bot }=\sqrt{D^{\pm}_{\bot}/c^2}$, $D^{\pm}_{\bot}=c^2k^2+\omega_{pe \pm}^{2}/\gamma_{0\pm}-\omega^2$, $\omega_{pe \pm}^{2}=e^2 n^{\pm} (1+m_e/m_i)/ \epsilon_0 m_e$, and $\gamma_{0\pm}=1/\sqrt{1-(v_0^\pm/c)^2}$. 
By evaluating the difference of the electric field derivatives across the shear interface, we obtain $\mathrm{I}.\delta\mathrm{E_0}=0$, where $\delta\mathrm{E_0}=(\delta E_y(0),\delta E_z(0))$, $I_{ij}=a_{ij}^+ k_{\bot}^+ + a_{ij}^- k_{\bot}^-$, $a^\pm_{11} =(\omega^2 -\omega^2_{pe \pm}/\gamma_{0\pm}) D^{\pm -1}_{\bot}$, $a^\pm_{12} = -(kv_0/\omega)(\omega^2_{pe \pm}/\gamma_{0\pm})D^{\pm-1}_{\bot}$ , $a^\pm_{21} = a^\pm_{12}$, and $a^\pm_{22} = -1 -(k^2c^2/\omega^2-1)(\omega^2_{pe \pm}/\gamma_{0\pm})(v_0^2/c^2) D^{\pm -1}_{\bot}$. We obtain the growth rate ${\rm Im}(\omega)$ for different $n_+/n_-$ ($n_{\rm jet}/n_{\rm out}$) by numerically solving
\begin{equation}
\label{eq:det_I}
\mathrm{det}(\mathrm{I}) =0
\end{equation}
Figure \ref{fig2}(a) shows the obtained growth rate of the MI or the imaginary part of the frequency ($\omega$) for different density contrasts. The growth rate is smaller for $n_+/n_-=2$ than for $n_+/n_-=1$. For comparison, in Figure \ref{fig2}(b) we show our simulation results and the difference in the evolution of the magnetic field energy, when the density ratio $n_{\rm jet}/n_{\rm out}$
varies from 1.25, to 5 (Runs 31-34, Table~\ref{tab:table1}). 
The saturation time is delayed as the ratio increases, which means that the initial growth rate is smaller for the larger $n_{\rm jet}/n_{\rm out}$. This is qualitatively consistent with Figure \ref{fig2}(a). We note that laser plasma experiments
\cite{ma20} have also shown that the saturation time is delayed as $n_{\rm PDG}/n_{\rm out}$ (in which PDG stands for Plasma Density Grating. ) increases. 
Runs 30-34 are performed to study the evolution of the jet for the different surrounding densities (Table~\ref{tab:table1}). Figure~\ref{fig2_2} are the magnetic field distributions for these runs and show the influence of the surrounding matter on the jet. 
As $n_{\rm jet}/n_{\rm out}$ increases from Run~30 to 34, the jet radius increases. 
In our simulations, the thermal velocity of the particles surrounding the jet is constant ($v_{\rm thr}=0.1\: c$). Thus, the decrease of $n_{\rm out}$ or the increase of $n_{\rm jet}/n_{\rm out}$ corresponds to the decrease of the external pressure of the jet. The difference in jet radius between Runs 30--34 indicates the importance of the external pressure for jet confinement. This effect has actually been discussed for astrophysical jets \cite{1989ApJ, gour18}.

\begin{figure}
\includegraphics[width=3.in]{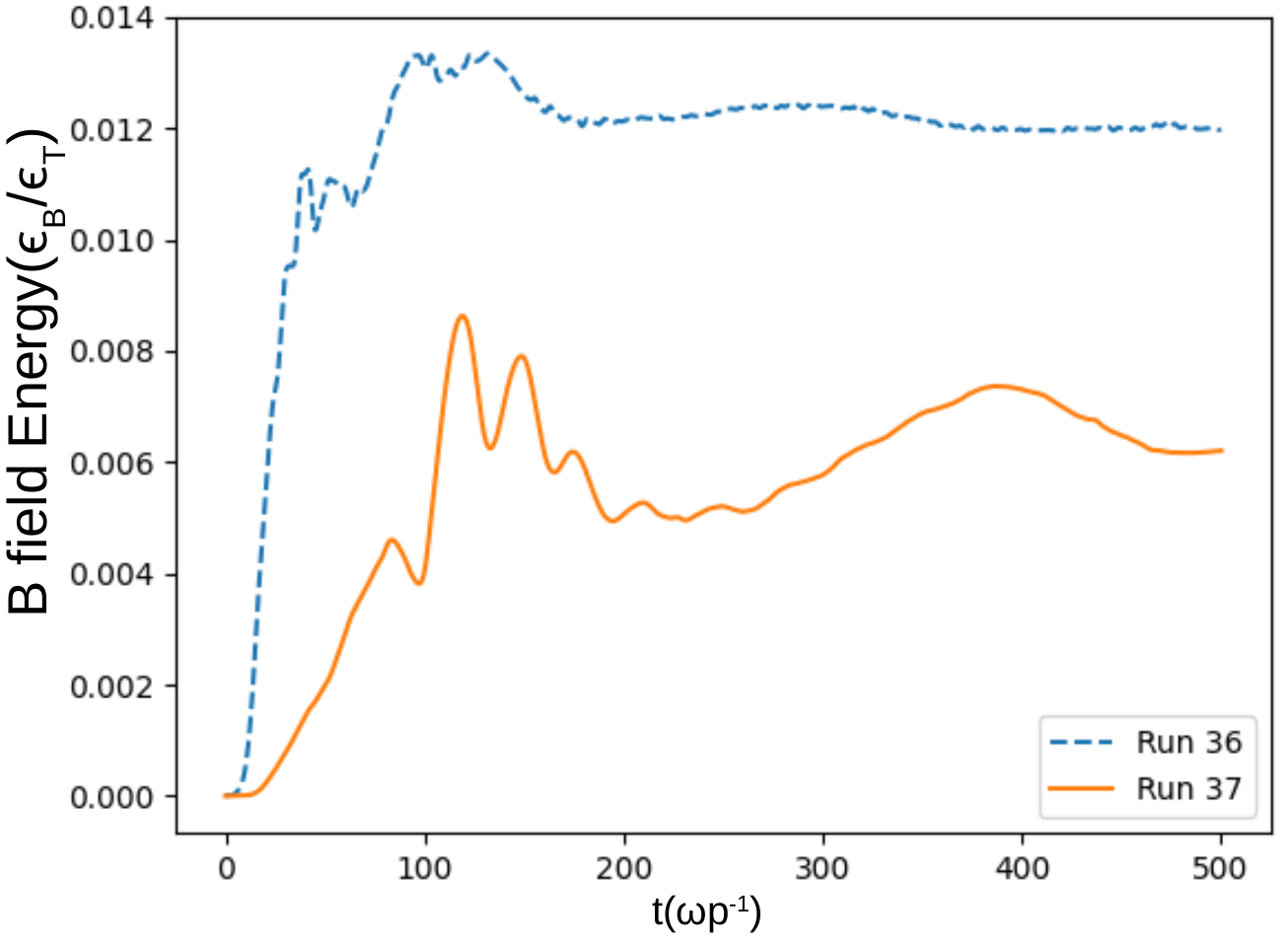}
\caption{\label{fig3} Evolution of magnetic field energy for different Lorentz factors ($\gamma_{0} =2$ for Run~36 and $\gamma_{0} =10$ for Run~37).  
}\end{figure}

\begin{figure*}
\includegraphics[width=7.in]{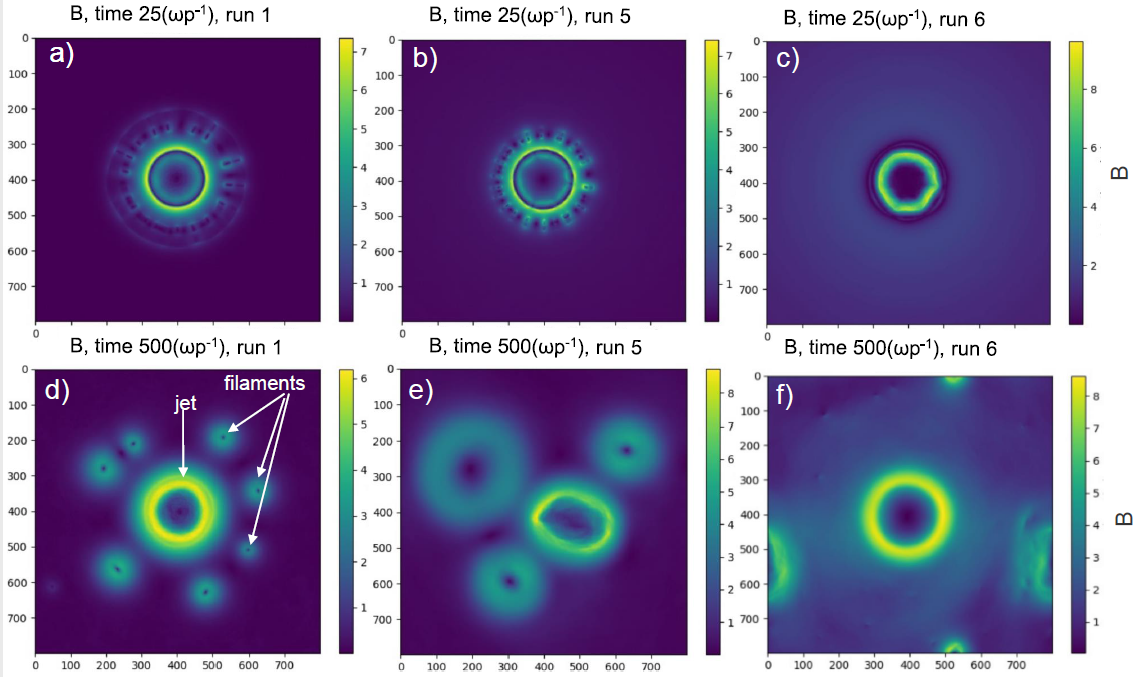}  
\caption{\label{fig4} Distribution of the magnetic field strength. The top refers to the results at $t=25$, while the bottom refers to those at $t=500$. From the left row to the right row the results for 
(a), (d) Run 1, (b), (e) Run 5 and (c), (f) Run 6
are shown. 
The initial values are $b_{0}=0$ for Run 1, $0.01$ for Run 5 and $0.05$ for Run 6. The Lorentz factor is $\gamma_0=2.3$ for all runs. In Run 6
, panel (f)
, the filaments are formed far away from the jet. We note that compared to Run~8 ($\gamma_0 =1.16$ and $b_0=0.01$, see Figure 1),
the initial jet velocity is larger and consequently the generated magnetic field is also larger. In panel (d) the jet and some filaments are marked. 
}
\end{figure*}

\subsection{The influence of Larger Lorentz factor for non-magnetized plasma}

For initially non-magnetized plasma, Alves et al.\cite{alves15} analytically studied the effect of the Lorentz factor on the growth rate of the MI for a plane geometry as shown in their Figure~1(b). They showed that the growth rate of the MI behaves as $\gamma_0^{-1/2}$ in the range $\gamma_0\gtrsim 2$, while it is an increasing function of $\gamma_0$ for $\gamma_0\lesssim 2$. To investigate whether their results can be applied to our jet geometry, we study the growth rate of the magnetic field energy
with different jet Lorentz factors of $\gamma_{0} = 2$ and $\gamma_{0} = 10$ (Runs 36 and 37).  We note that we give the thermal velocity in our simulations ($v_{\rm out}=v_{\rm thr}=0.1c$), and the influence is relatively strong when there is no initial magnetic field. Therefore, we make a comparison only for $\gamma_{0} \geq 2$, since the thermal velocity can be ignored. Figure~\ref{fig3} shows the results. Since the initial magnetic field is set to zero, the magnetic field is generated only by the motion of the jet particles. For comparison, the curve for $\gamma_{0} =2$ (Run~36) is multiplied by 10. 
As can be seen, the curve for $\gamma_{0} =2$ is steeper than that for $\gamma_{0} =10$ (Run~37) during the initial growth phase ($t\lesssim 30$ for Run~36 and $t\lesssim 100$ for Run~37). This shows that a very large Lorentz factor prevents rapid growth of magnetic fields. In fact, we calculate the growth rates and find that $\Gamma=0.063$ for Run 37 and $\Gamma=0.15$ for Run 36 and the ratio is $0.15/0.063 =2.38$ which is consistent with the theory $(\Gamma\propto \gamma_0^{-1/2}$ or $\sqrt{10}/ \sqrt{2}=2.24)$.

\begin{figure*}
\includegraphics[width=7.in]{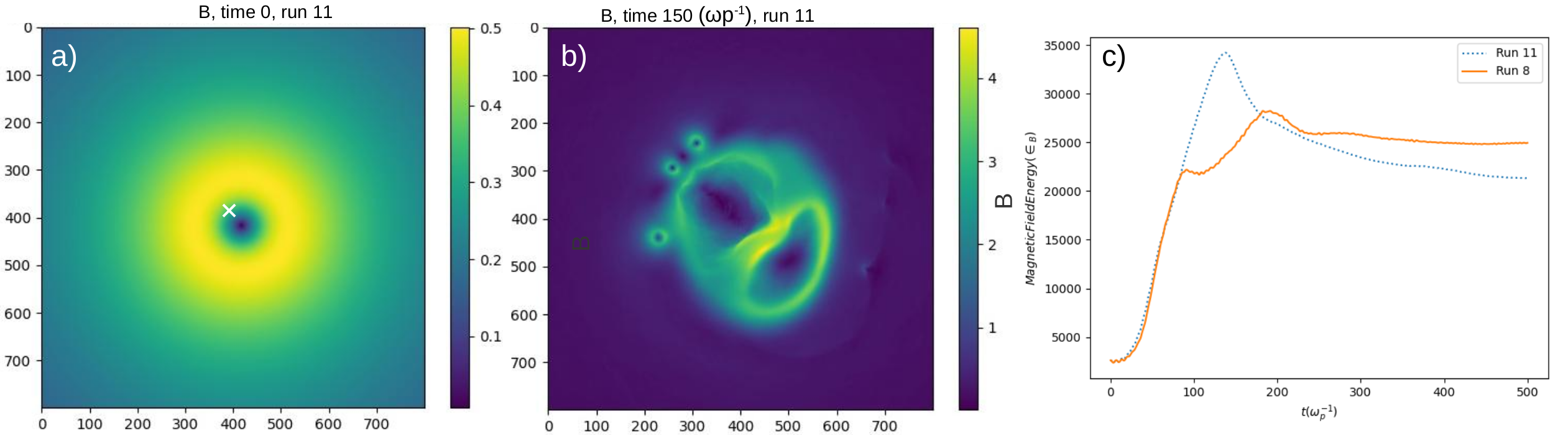}  
\caption{\label{fig5} Effect of the off-center initial magnetic field on the instability. (a) Magnetic field distribution for Run 11 at $t=0$. The field is toroidal and surrounds the jet cross section, but is shifted +20 in the $x$, $y$ directions. The cross represents the center of the jet. (b) Magnetic field distribution for Run 11 at $t\sim 150$. No filaments are formed in the lower right corner of the jet, in contrast to the
filaments in the upper left corner. (c) Evolution of the total magnetic field energy for Runs 8 (fiducial model) and 11.}

\end{figure*}

\begin{figure*}
\includegraphics[width=7.in]{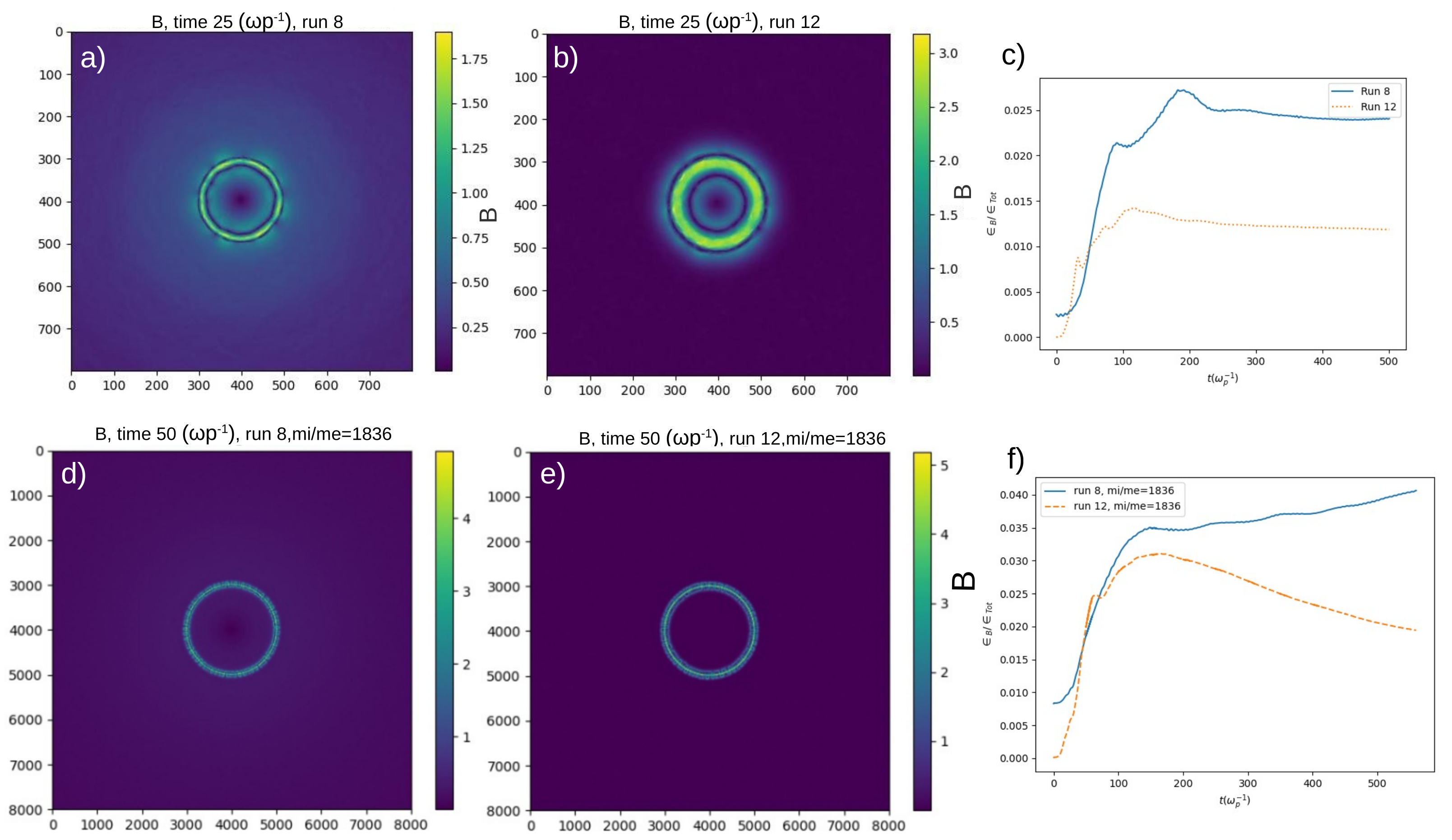}  
\caption{\label{fig6} Effect of the initial magnetic field topology on the instability. Magnetic field distribution at time $t=25$ for the case of (a) toroidal initial magnetic field (Run 8; $\gamma_0=1.16$ and $b_{0}=0.01$) and (b) axial initial magnetic field (Run 12; $\gamma_0=1.16$ and $b_{0}=0.01$). The axial initial magnetic field generates more magnetic field at later times and also stabilizes the jet. (c) Evolution of the total magnetic field energy 
divided by the total energy (kinetic plus electromagnetic 
for Runs 8 and 12. The growth rate at $t\lesssim 25$ for Run 12 is higher than that for Run 8. 
Panels (d), (e), (f) are the same as panels (a), (b), (c), but for the larger size of $8000\times 8000$ and a mass ratio of 1836. The simulation time is $t=50$ for panels (d) and (e).
}
\end{figure*}

\subsection{Initially magnetized plasma}  

\subsubsection{Dependence of filament position on toroidal magnetic field}

We consider the effect of the initial magnetic field on the evolution of the MI when the field is toroidal.
We run three simulations with $b_{0} =0, 0.01$
and $0.05$ (Runs 1, 5, and 6). Figures~\ref{fig4}(a)-(c) show that the magnetic fields have been amplified by the time $t=25$; their maximum strengths are 7.5, 7.5, and 9.0, respectively. Figures~\ref{fig4}(d)-(f) show that by $t=500$, magnetic fields have also been created in the form of filaments around the jets by the MI (see Figure~\ref{fig4}(d) for the definition). The positions of the filaments are further away from the jet for larger $b_0$ because the initial field pushes the filaments out. The results complement a previous MHD study\cite{mizuno15}.

\subsubsection{Off center effects for toroidal magnetic field}  %
To see the effect of the initial magnetic field configuration on the jet, we shifted the center of the toroidal initial
magnetic field in Run 11, for which the other parameters are the same as those for Run 8. In Figure \ref{fig5}(a), the center is moved 20 grids to the lower right of the simulation box. At $t=150$, the side with the larger magnetic field (bottom right) does not have filaments and is more stable than the opposite side (top left) as is shown in Figure \ref{fig5}(b). As a result, the maximum strength of the magnetic field is larger in Run 11 than in Run 8 (Figure \ref{fig5}(c)).

\subsubsection{Axial magnetic field}
\label{sec:axial}

 We also consider the case where the initial magnetic field is axial and oriented in the $z$ direction to see how the initial field topology affects the evolution of the magnetic field. In Figure \ref{fig6} we compare the results for the axial field (Run 12) with those for the toroidal field (Run 8) at an early stage of evolution ($t=25$). The field is generated more effectively for Run 12 (Figure \ref{fig6}(b)) than for Run 8 (Figure \ref{fig6}(a)). 
In fact, the maximum value of the field in Figure \ref{fig6}(b) is 3.2, while
that in Figure \ref{fig6}(a) is 1.9. 
In addition, the thickness of the magnetized ring-like region is larger in Run 12 (Figure \ref{fig6}(b)), which stabilizes the jet. Figure \ref{fig6}(c) shows that the growth rate at $t\lesssim 25$ for the run with the axial initial magnetic field (Run 12) is higher than that of the toroidal case (Run 8). 
Interestingly, this is in agreement with the results of MHD simulations, although MHD simulations cannot reveal the initial field growth in detail. For example, Mizuno et al.\cite{mizuno15} studied the growth of magnetic fields for three different initial fields (helical, toroidal, and axial) using 3D MHD simulations.
They showed that the axial field leads to the most efficient growth of the magnetic field, and the next is the helical field, which includes an axial component.

Figures \ref{fig6}(d)--(f) are the same as Figures \ref{fig6}(a)--(c) but for a larger size of $8000\times 8000$ with $m_i/m_e=1836$. 
At $t=50$, the total magnetic field is stronger for Run 12 (Figure \ref{fig6}(e)) than for Run 8 (Figure \ref{fig6}(d)). Also, the initial growth rate of Run 12 is higher than that of Run 8 (Figure \ref{fig6}(f)). The results of the larger size with realistic mass ratio are at least qualitatively consistent with those of the $800 \times 800$ size used in this paper.

\begin{figure}
\includegraphics[width=3.in]{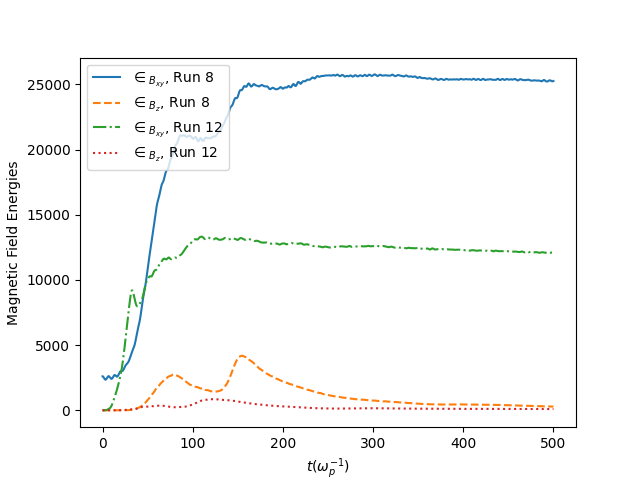}  
\caption{\label{fig1_1}Axial and toroidal components of the magnetic field energy growth/saturation for different initial magnetic fields. Run 8 is for the initial toroidal magnetic field: $(b_{0}, \gamma_0)=(0.01, 1.16)$. Runs 12 is for the initial axial magnetic field: $(b_{z0}, \gamma_0)=(0.01, 1.16)$.}
\end{figure}

Magnetic field energy growth for axial $\in_{B_z}$ and toroidal $\in_{B_{xy}}$ components is shown in Figure \ref{fig1_1}. It shows that both the initial axial (Run~12) and toroidal (Run~8) field jets generate an axial $z$ component of the magnetic field. The initial magnetic field can be ignored in the figure. The $z$ component develops initially, and the value of the axial component is much smaller than that of the toroidal component. For example, in Run~8, the $z$ component is $\in_{B_z} \sim 4200$ at $t\sim 160$, which is much smaller than the toroidal component ($\in_{B_{xy}}\sim 25000$ at $t\sim 160$). Later, the $z$ component decays to small values. This may be consistent with the magnetic structure in the jets being helical\cite{gabuzda19a, gabuzda19}. In fact, Gabudza [2019] pointed out that due to the faster decay of the axial component of the magnetic field, the toroidal component should become dominant.\cite{gabuzda19a} Figure \ref{fig1_1} confirms this behavior in our simulations.
\subsubsection{Sensitivity to initial parameters}
\begin{figure}
\includegraphics[width=3.in]{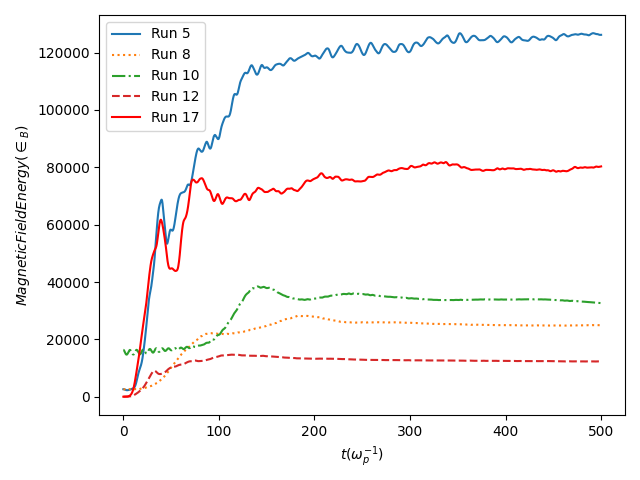}  
\caption{\label{fig7}Magnetic field energy growth/saturation for different initial magnetic fields and jet velocities. Runs 5, 8, and 10 are for toroidal field: $(b_{0}, \gamma_0)=(0.01, 2.3),(0.01, 1.16),(0.025, 1.16)$. Runs 12, 13, and 17 are for axial field: $(b_{0}, \gamma_0)=(0.01, 1.16), (0.025, 1.16), (0.01, 2.3)$. The evolution of Run 13 is almost the same as that of Run 12, so it is not shown.}
\end{figure}

Figure~\ref{fig7} shows the evolution of the magnetic field energy for different initial parameters. Runs 5, 8, and 10 have the toroidal initial field. For Run 10, the initial field strength $b_0=0.025$ is 2.5 times larger than that of our fiducial model or Run 8 ($b_0=0.01$). However, the saturated magnetic field energy for Run 10 at $t\sim 400$ is only 1.35 times larger than that for Run 8. On the other hand, the Lorentz factor for Run 5 ($\gamma_0=2.3$) is about two times larger than that for Run 8 ($\gamma_0=1.16$). The saturated field for Run 5 is 5.1 times larger than Run 8 at $t\sim 400$. 
These show that the evolution of the magnetic field is more sensitive to the Lorentz factor than to the initial magnetic field strength. 

  The same can be said for the cases where the initial field is axial. For example, $b_0=0.025$ for Run~13 while $b_0=0.01$ for Run~12. However, Run~13 has nearly the same profile as Run~12 in Figure~\ref{fig7}.
On the other hand, the Lorentz factor of $\gamma_0=2.3$ for Run 17 is about two times larger than $\gamma_0=1.16$ for Run 12.
The saturated field for the former is 6.4 times larger than the latter ($t\sim 400$ in Figure~\ref{fig7}). 

\section{\label{sec:level1_3} Summary}
We have studied the generation of the magnetic field around a jet using a two-dimensional particle code. We investigated the evolution of the field in the cross section of the jet.
We focused on the effects of different initial parameters such as the density of the jet environment, the Lorentz factor, and the magnetic field strength and structures (toroidal and axial). We have
compared our results with those of previous studies. 

Our simulations demonstrate that, in general, magnetic fields rapidly develop due to plasma instabilities. Currents and their corresponding magnetic filaments form around the jet and merge together, resulting in amplified magnetic fields. Nevertheless, the Alfv\'en current limit eventually causes saturation of the field. 
For plasma without initial magnetization, we show that if the matter density of the jet environment is lower, there is a delay in the growth of the magnetic field. Additionally, if the Lorentz factor is higher ($\gtrsim 2$), it prevents quick growth of the magnetic fields.
When a toroidal, non-zero magnetic field is present initially, it influences the instability at the boundary of the jet. A greater initial field pushes the filaments away from the jet. 
If the initial magnetic field is off-center, filaments are not generated on the stronger side, but they are generated on the weaker side. 
We also found that an axial initial magnetic field creates a thick magnetic region around the jet, which stabilizes the jet. As a result, a jet with an axial initial magnetic field is more stable than one with a troidal field.
\\As noted in Section II, the difference in physical scales does not allow us to quantitatively compare our simulation results with observations of real astrophysical jets. 
Nevertheless, our simulations suggest the jet confinement by external pressure (Section III.C) and the development of helical magnetic field (Section III.E.3), which may be consistent with the observations of astrophysical jets.

\begin{acknowledgments}
The authors wish to thank Victor Decyk and Luis Silva for their fruitful discussion about the simulation and analysis. Numerical computations were [in part] carried out on Cray XC50 at Center for Computational Astrophysics, National Astronomical Observatory of Japan. This work was supported by JSPS KAKENHI Grant Number JP22H00158, JP22H01268, JP22K03624, JP23H04899 (Y.F.).
\end{acknowledgments}

\section*{Data Availability Statement}

The data that support the findings of this study are available from the corresponding author upon reasonable request.

\nocite{*}
\bibliography{aipsamp}


\end{document}